\documentclass[11pt]{article}
 \usepackage{mathrsfs}
 \usepackage{bbm}
 \usepackage{amsthm}
\newcommand{\beq}{\begin{equation}}
\newcommand{\eeq}{\end{equation}}
\newcommand{\ba}{\begin{array}}
\newcommand{\ea}{\end{array}}

\newtheorem{theorem}{Theorem}[section]
\newtheorem{lemma}[theorem]{Lemma}

\theoremstyle{definition}

\theoremstyle{remark}

\theoremstyle{algorithm}
\newtheorem{algorithm}[theorem]{Algorithm}

\theoremstyle{corollary}

\theoremstyle{example}

\usepackage{amssymb}
\usepackage{epsfig}
\usepackage{times}
\title{Distributed Bayesian Detection Under Unknown Observation
Statistics}

\author{Xiaojing Shen, Member, IEEE, Pramod K. Varshney, Fellow,
IEEE, Yunmin Zhu
\thanks{This work was supported  in part by  U.S. Air Force Office of
Scientific Research (AFOSR) under Grants FA9550-10-1-0263 and
FA9550-10-1-0458 and in part by the NNSF of China (\# 61004138 and
61273074).} 
\thanks{Xiaojing Shen and Pramod K. Varshney (corresponding author, varshney@syr.edu) are with the Department of Electrical
Engineering and Computer Science, Syracuse University, NY, 13244,
USA. Yunmin Zhu is  with Department of Mathematics, Sichuan
University, Chengdu, Sichuan 610064, China.  Xiaojing Shen
(shenxj@scu.edu.cn) is on leave from Department of Mathematics,
Sichuan University, Chengdu, Sichuan 610064, China.}}

\begin{document}
\maketitle
\begin{abstract}
In this paper, distributed Bayesian detection problems with unknown
prior probabilities of hypotheses are considered. The sensors obtain
observations which are conditionally dependent across sensors and
their probability density functions (pdf) are not exactly known. The
observations are quantized and are sent to the fusion center. The
fusion center fuses the current quantized observations and makes a
final decision. It also designs (updated) quantizers to be used at
the sensors and the fusion rule based on all previous quantized
observations. Information regarding updated quantizers is sent back
to the sensors for use at the next time.  In this paper, the
conditional joint pdf is represented in a parametric form by using
the copula framework. The unknown parameters include dependence
parameters and marginal parameters. Maximum likelihood estimation
(MLE) with feedback based on quantized data is proposed to estimate
the unknown parameters. These estimates are iteratively used to
refine the quantizers and the fusion rule to improve distributed
detection performance by using feedback. Numerical examples show
that the new detection method based on MLE with feedback is much
better than the usual detection method based on the assumption of
conditionally independent observations.
\end{abstract}

\noindent{\bf keywords:}  Bayesian detection, copula-based
dependence modeling, copulas, maximum likelihood estimation,
distributed detection, information fusion

\section{Introduction}
Distributed detection has received considerable attention over the
last few decades
\cite{Varshney97,Vismanathan-Varshney97,Blum-Kassam-Poor97,Zhu-Zhou-Shen-Song-Luo12,Veeravalli-Varshney12}.
The Bayesian formulation of distributed detection was first
considered by Tenney and Sandell \cite{Tenney-Sandell81} for
parallel sensor network structures. For conditionally independent
sensor observations, they proved that the optimal decision rules at
the sensors are likelihood ratio (LR) quantizers. The optimal
thresholds to quantize LR at individual sensors can be determined by
solving a set of coupled nonlinear equations. When the quantizers
are fixed, Chair and Varshney in \cite{Chair-Varshney86} derived an
optimum fusion rule, once again based on the LR test. Over the past
several years, many excellent results on distributed detection based
on the assumption of conditionally independent sensor observations
have been derived that are available in \cite{Varshney97} and
references therein. The emerging wireless sensor networks
\cite{Chen-Tong-Varshney06} motivated the optimality of LR
quantizers to be extended to non-ideal detection systems where
sensor outputs are to be communicated through noisy, possibly
coupled channels to the fusion center
\cite{Chen-Willett05,Chen-Chen-Varshney09}.

When sensor observations are dependent, Tsitsiklis and Athans
\cite{Tsitsiklis-Athans85} provided a rigorous mathematical analysis
demonstrating the computational difficulty in obtaining the optimum
quantizers. Some progress has been made for the dependent
observations case (see
\cite{Blum-Kassam92,Chen-Papamarcou95,Willett-Swaszek-Blum00,Tang-Pattipati-Kleinman92}).
For example, difficulties encountered when dealing with dependent
observations were discussed in \cite{Willett-Swaszek-Blum00}. In
\cite{Zhu-Blum-Luo-Wang00}, for  distributed dependent observations
and a fixed fusion rule, the authors proposed a computationally
efficient iterative algorithm for computing a discrete approximation
of the optimal quantizers. The finite-step convergence of this
algorithm was proved. By combining the methods proposed in
\cite{Chair-Varshney86} and \cite{Zhu-Blum-Luo-Wang00},
 an efficient algorithm to
simultaneously search for the optimal fusion rule and the optimal
quantizers was derived in \cite{Shen-Zhu-He-You11}. Recently, the
authors of \cite{Chen-Chen-Varshney12} introduced a new framework
for distributed detection with conditionally dependent observations.
The new framework can identify several classes of problems with
dependent observations whose optimal quantizers resemble the ones
for the independent case. In addition, copula-based distributed
Neyman-Pearson detection and hypothesis testing using heterogeneous
dependent data have been proposed in
\cite{Sundaresan-Varshney-Rao11,Iyengar-Varshney-Damarla11}. The
copula based approach provides a systematic and elegant approach to
characterize dependence and obtain decision rules at the sensors and
the fusion center.

\begin{figure}[h]\label{fig_0}
\vbox to 8.5cm{\vfill \hbox to \hsize{\hfill
\scalebox{0.5}[0.5]{\includegraphics{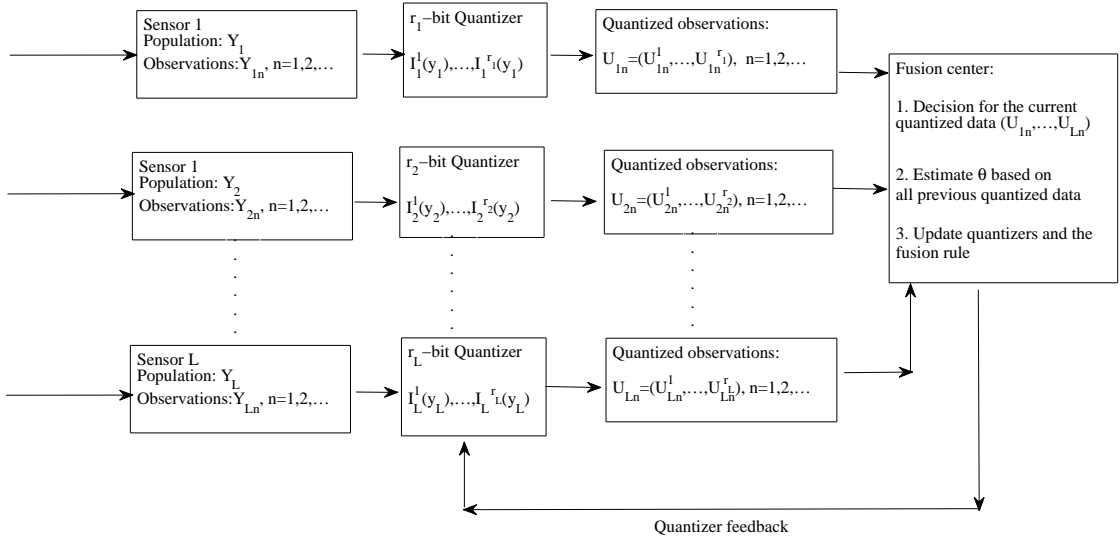}}
\hfill}\vfill}
\caption{Distributed detection system with feedback of quantizer
information}
\end{figure}

In all previous  studies on distributed Bayesian detection with
dependent observations
\cite{Tsitsiklis-Athans85,Blum-Kassam92,Chen-Papamarcou95,Willett-Swaszek-Blum00,Tang-Pattipati-Kleinman92,Zhu-Blum-Luo-Wang00,Shen-Zhu-He-You11,Chen-Chen-Varshney12},
the conditionally dependent joint pdfs of the sensor observations
are assumed known.  When the dependence among sensors is unknown,
the usual approach is to ignore dependence and assume that the
sensor observations are independent. The focus of this paper is
distributed Bayesian detection in the context of unknown
conditionally dependent pdfs. We also assume that prior
probabilities of the hypotheses are unknown. The specific scenario
(see Figure \ref{fig_0}) is that the sensors obtain observations
which are conditionally dependent across sensors. The observations
are quantized and are sent to the fusion center. The fusion center
fuses the current quantized observations and makes a final decision.
It designs (updated) quantizers to be used at the sensors and the
fusion rule based on all previous quantized observations.
Information regarding updated quantizers is sent back to the sensors
for use at the next time.


In \cite{Shen-Zhu-He-You11}, we have presented an iterative
algorithm to design the quantizers at the sensors and the fusion
center for Bayesian distributed detection. In this paper, we extend
the work in \cite{Shen-Zhu-He-You11} to the situation when the prior
probabilities and the joint pdf of conditionally dependent
observations are not known. The conditional joint pdf is represented
in a parametric form by using the copula framework. The unknown
parameters include dependence parameters and marginal parameters
(parameters corresponding to marginal pdfs). MLE with feedback based
on quantized data is proposed to estimate the unknown parameters.
Its asymptotic efficiency can be guaranteed by employing the result
that we have developed in \cite{Shen-Varshney-Zhu12} with an
asymptotic variance which is equal to the inverse of a convex linear
combination of Fisher information matrices based on $J$ groups of
different feedback quantizers.
These estimates are iteratively used to refine the quantizers and
the fusion rule to improve distributed detection performance by
using feedback. Numerical examples show that the new  detection
method based on MLE with feedback is much better than the usual
detection method based on the assumption of conditionally
independent observations. Better detection performance can be
obtained by increasing the number of feedbacks and the number of
observations during each estimation step.

The rest of the paper is organized as follows. Problem formulation
is given in Section \ref{sec_2}. In Section \ref{sec_3},
copula-based parametric pdfs  are constructed.  MLE with feedback
based on quantized data is also proposed. In Section \ref{sec_4}, an
efficient distributed detection algorithm  with unknown pdfs and
unknown joint prior probabilities is presented based on quantized
observations and different updated quantizers. In Section
\ref{sec_5}, numerical examples are given that exhibit the superior
performance of our approach. In Section \ref{sec_6}, concluding
remarks are provided.

\section{Problem Formulation}\label{sec_2}
An $L$-sensor Bayes detection system with two hypotheses $H_0$ and
$H_1$ is considered.  A parallel architecture with feedback is
assumed (see Figure \ref{fig_0}). Each sensor acquires observations
$Y_i, i=1,\ldots,L$ whose dimension is assumed to be one for
notational simplicity in this paper. The case of high dimensions can
be similarly considered.
The $i$-th sensor quantizes the observation vector to $r_i$ bits
($r_i\geq1$) by $r_i$ measurable indicator quantization functions:
\begin{eqnarray}%
\label{Eq2_4} && I_i^1(y_i):~y_i\in\mathbb{R}\rightarrow
\{0,1\};~\ldots~;~ I_i^{r_i}(y_i):~y_i\in\mathbb{R}\rightarrow
\{0,1\},
\end{eqnarray}
for $i=1,\ldots, L$. Here, each binary quantizer $I_i^{t}(y_i)$
partitions the space into two regions that could be continuous
regions or union of discontinuous regions. Moreover, we denote the
quantization functions by
\begin{eqnarray}%
\label{Eq2_5}&&I(y|r)\triangleq (I_1(y_1)',\ldots,I_L(y_L)')' \in
\mathbb{R}^r,
\end{eqnarray}
where
\begin{eqnarray}%
\label{Eq2_6}&&I_i(y_i)\triangleq(I_i^1(y_i),\ldots,I_i^{r_i}(y_i))',~
i=1,\ldots,L,
\end{eqnarray}
and $r=\sum_{i=1}^Lr_i$ is the total number of bits available to
transmit observations from the sensors to the fusion center. Once
the $r_i$-bit binary quantized measurements $I_i(Y_i)$ are generated
at sensor $i$, $i=1,\ldots,L$, they are transmitted to the fusion
center. The fusion center then makes a final decision 0/1 based on
$I_i(Y_i), i=1,\ldots,L$ using a fusion function $F(u_1,\ldots,u_L),
u_i=0/1, i=1,\ldots,L$, i.e.,
\begin{eqnarray}%
F(I_1(Y_1),\ldots,I_L(Y_L))=0/1.
\end{eqnarray}

If the prior probabilities of hypotheses and the conditional pdfs
$p(y_1,\ldots,y_L|H_j),j=0,1$ are known, the goal of the distributed
Bayesian detection system is to design a set of optimal sensor
quantizers $I_1(y_1)$, $\ldots$, $I_L(y_L)$ and an optimum fusion
rule $F(u_1,\ldots,u_L)$ such that the following Bayes cost
functional is as small as possible.
\begin{eqnarray}
\nonumber &&C(I_1(y_1),\ldots,I_L(y_L);F)\\
\nonumber&\triangleq&C_{00}P_0P(F(I_1(Y_1),\ldots,I_L(Y_L))=0|H_0)+C_{01}P_1P(F(I_1(Y_1),\ldots,I_L(Y_L))=0|H_1)\\
\label{Eq4_3}
&&+C_{10}P_0P(F(I_1(Y_1),\ldots,I_L(Y_L))=1|H_0)+C_{11}P_1P(F(I_1(Y_1),\ldots,I_L(Y_L))=1|H_1),
\end{eqnarray}
where $C_{ij}$ are cost coefficients; $P_0$ and $P_1$ are the prior
probabilities for the hypotheses $H_0$ and $H_1$; $P($
$F(I_1(Y_1),\ldots,I_L(Y_L))=i |H_j)$ is the probability that the
fusion center decides in favor of hypothesis $i$ given hypothesis
$H_j$, $i,j=0,1$, which can be computed based on
$p(y_1,\ldots,y_L|H_j),j=0,1$ (see, e.g., \cite{Varshney97}).

However, in distributed detection systems with limited bandwidth,
the joint conditional pdfs are hard to obtain by traditional pdf
estimation methods based on raw observations. In some situations,
the prior probabilities of $H_0$ and $H_1$ are also unknown.
Quantized observations and initial quantizers is all the information
available at the fusion center. The specific scenario considered
here is that the sensors obtain observations which are conditionally
dependent across sensors. The observations are quantized and are
sent to the fusion center. The fusion center fuses the current
quantized observations and makes a final decision. It also designs
(updated) quantizers to be used at the sensors and the fusion rule
based on all previous quantized observations. Information regarding
updated quantizers is sent back to the sensors for use at the next
time.

In summary, when the prior probabilities and conditionally dependent
pdfs are unknown, the fusion center faces the problem of how to make
a decision for the current set of quantized observations and the
problem of how to improve the detection performance by using all
previous quantized observations in time.

The first problem is how to design the quantizers and the fusion
rule, which is a static problem when the prior probabilities and the
conditional pdfs are known. Previously developed methods (see, e.g.,
\cite{Chair-Varshney86,Tang-Pattipati-Kleinman92,
Zhu-Blum-Luo-Wang00,Shen-Zhu-He-You11,Chen-Chen-Varshney12}) can be
used to solve the problem. Thus, we concentrate on how to
update/estimate the unknown prior probabilities and conditionally
dependent pdfs using all the previous quantized data. Note that the
problem relies on the updated quantizers that are fed back to the
sensors, which results in quantized observations not having
identical distributions temporally.

\section{Copula-based Maximum Likelihood Estimation with Feedback}\label{sec_3}
\subsection{Copula-based dependence modeling}
In distributed detection  with dependent observations, the
performance of the detection system depends on the exploitation of
dependence among sensor measurements. In most previous works, when
dependence is not known, independence is usually assumed  across
sensors for simplicity. Here, we will model dependence by parametric
copulas. Actually, copula is a distribution function whose
one-dimensional marginals are uniform.


\begin{lemma}
(Sklar's Theorem, see \cite{Sklar59} or \cite{Nelsen99}) Let $F$ be
an $L$-dimensional distribution function with marginals $F_1,
F_2,\ldots, F_L$. Then there exists an $L$-copula
$C(v_1,v_2,\ldots,v_L)$, $v_i\in[0, 1], i=1,\ldots,L$ such that for
all $(y_1,\ldots, y_L)$ in $\mathbb{R}^L$ ,
\begin{eqnarray}\label{Eq4_01}
F(y_1,y_2,\ldots, y_L)=C(F_1(y_1),F_2(y_2),\ldots,F_L(y_L)).
\end{eqnarray}
 If $F_1,F_2, \ldots, F_L$ are all continuous, then $C$ is unique;
otherwise, C is uniquely determined on $Ran~ F_1\times Ran~
F_2\times Ran~ F_L$. Conversely, if $C$ is an $L$-copula and
$F_1,F_2, \ldots, F_L$ are distribution functions, then the function
$F$ defined by (\ref{Eq4_01}) is an $L$-dimensional distribution
function with marginals $F_1,F_2, \ldots, F_L$.
\end{lemma}

From Sklar's Theorem, the joint pdf is equivalent to
\begin{eqnarray}
\label{Eq4_1}
p(y_1,y_2,\ldots,y_L)=c(F_1(y_1),F_2(y_2),\ldots,F_L(y_L))\prod_{i=1}^Lp_i(y_i),
\end{eqnarray}
where $p_i(y_i)$ and $F_i(y_i), i=1,\ldots,L$ are marginal pdfs and
distribution functions of continuous random variables respectively;
$c(v_1,v_2,\ldots,v_L)$,  $v_i\in[0,1], i=1,\ldots,L$ is the copula
density function,
\begin{eqnarray}
\label{Eq4_001}
c(v_1,v_2,\ldots,v_L)=\frac{\partial^LC(v_1,v_2,\ldots,v_L)}{\partial
v_1,\ldots,\partial v_L}.
\end{eqnarray}

If measurements are conditionally independent, then
$p(y_1,y_2,\ldots,y_L)=\Pi_{i=1}^Lp_i(y_i)$ and $c(v_1,v_2,\ldots,$
$v_L)\equiv1$. When the measurements are dependent,
$c(v_1,v_2,\ldots,v_L) \neq 1$ and all dependence information among
measurements is contained in $c(v_1,v_2,\ldots,v_L)$. Thus, copula
framework allows us flexibly represent the dependence of
observations at the sensors by $c(v_1,v_2,\ldots,v_L)$ which is
independent of the marginal pdfs so that marginal pdfs can be
arbitrary pdfs of continuous random variables and not be limited to
Gaussian pdfs.
Since the nonparametric copula estimation methods require heavy
computations and cannot easily use the knowledge of quantizers, we
concentrate on the parametric estimation of $c(v_1,v_2,\ldots,v_L)$.
Thus, the parametric structure
$c(v_1,v_2,\ldots,v_L|H_j,\theta_{0j})$ and
$p_i(y_i|H_j,\theta_{ij})$ are assumed known under hypothesis $H_j$,
and then we have the joint pdf under hypothesis $H_j$ as follows
\begin{eqnarray}
\label{Eq4_2}
p(y_1,y_2,\ldots,y_L|H_j,\theta_j)=c(F_1(y_1|H_j),F_2(y_2|H_j),\ldots,F_L(y_L|H_j)|H_j,\theta_{0j})\prod_{i=1}^Lp_i(y_i|H_j,\theta_{ij}),
\end{eqnarray}
where
$\theta_j\triangleq[\theta_{0j},\theta_{1j},\ldots,\theta_{Lj}]$ is
the parameter vector to be estimated; $\theta_{0j}$ is the
dependence parameter and $\theta_{1j},\ldots,\theta_{Lj}$ are the
marginal parameters under hypothesis $j=0, 1$. There exist many
parametric structures of copula density
$c(v_1,v_2,\ldots,v_L|H_j,\theta_{0j})$ such as Clayton copula,
Gumbel copula, Frank copula and Gauss copula, $t$ copula etc. (see,
e.g., \cite{Nelsen99}). The ``best" copula model can be selected by
criteria such as Akaike information criterion (AIC), AIC with a
correction (AIC$_c$) and Bayesian information criterion (BIC) etc (
see, e.g., \cite{Burnham-Anderson02}). Here, we assume that the
copula model has been determined but its parameters are not known.


\subsection{Maximum likelihood estimation of unknown prior probabilities and parameters of pdfs  with quantized observations}
The observation population of the $i$-th sensor is denoted $Y_i$,
$i=1,\ldots,L$. The observation samples of $Y_i$ may be from $H_0$
or $H_1$. The joint observation population is denoted by
$Y\triangleq(Y_1',\ldots,Y_L')'$ which has the following family of
joint pdf:
\begin{eqnarray}%
\label{Eq2_1} \{p(y_1,\ldots,y_L|\theta)\}_{\theta\in\Theta\subseteq
\mathbb{R}^k},
\end{eqnarray}
where  $\theta$ is the unknown $k$-dimensional deterministic
parameter vector which may include marginal parameters and
dependence parameters. Note that the conditionally joint pdf under
hypothesis $H_j$, $p(y_1,y_2,\ldots,y_L$ $|H_j,\theta_j)$ can be
constructed by (\ref{Eq4_2}) where
$\theta_j=[\theta_{0j},\theta_{1j},\ldots,\theta_{Lj}], j=0,1$ are
parameter vectors and the prior probabilities of $H_0$ and $H_1$ are
$P_0$ and $P_1=1-P_0$ respectively. Thus, we have
\begin{eqnarray}%
\label{Eq2_1} p(y_1,\ldots,y_L|\theta)&=&
P_0p(y_1,\ldots,y_L|H_0,\theta_0)+(1-P_0)p(y_1,\ldots,y_L|H_1,\theta_1),
\end{eqnarray}
where $\theta\triangleq[P_0,\theta_0,\theta_1]$ is the parameter
vector to be estimated. Note that $\theta_0$ and $\theta_1$
themselves are vectors. The true parameter vector (the clairvoyant
case) is denoted by $\theta^*$.

 Let $N$ independently and identically distributed
(i.i.d.) temporal sensor observation samples and joint observation
samples be
\begin{eqnarray}%
\label{Eq2_2}\vec{Y}_i&=&(Y_{i1},\ldots,Y_{iN}),~ i=1,\ldots,L;\\[3mm]
\label{Eq2_3} \vec{Y}&=&(\vec{Y}_1',\ldots,\vec{Y}_L')'.
\end{eqnarray}
Moreover, based on the definition of quantizers
(\ref{Eq2_4})--(\ref{Eq2_6}), we define the quantized sensor
observation samples and the joint quantized  observation samples as
follows
\begin{eqnarray}%
\label{Eq3_4} \vec{U}&\triangleq&(\vec{U}_{1}',\ldots,\vec{U}_{L}')',\\[3mm]
\label{Eq3_3} \vec{U}_{i}&\triangleq&(U_{i1},\ldots,U_{iN})',~
i=1,\ldots,L,\\[3mm]
\label{Eq3_2}
U_{in}&\triangleq&(U_{in}^1,\ldots,U_{in}^{r_i}),~~n=1,\ldots,N,\\[3mm]
\label{Eq3_1} U_{in}^1&\triangleq&
I_i^1(Y_{in}),\ldots,U_{in}^{r_i}~\triangleq~I_i^{r_i}(Y_{in}),
\end{eqnarray}
where $\vec{U}$ is the joint quantized observation samples. We
denote the quantized observation population by $U\triangleq I(Y|r)$
$=(I_1(Y_1),$ $\ldots$, $I_L(Y_L))'$, we know that $U$ has a
\emph{discrete/categorical} distribution. Based on the pdf of $Y$
and quantizers $I(y|r)$, the probability mass function (pmf) of the
quantized observation population $U$ is
\begin{eqnarray}
\label{Eq3_5} f_U(u_{1},u_{2},\ldots,u_{L}|\theta)=P_{u_{1},u_{2},\ldots,u_{L}}~~  \mbox{for}~~U=(u_{1},u_{2},\ldots,u_{L}),~~
\end{eqnarray}
where
\begin{eqnarray}
\nonumber (u_{1},u_{2},\ldots,u_{L})&\in& \mathcal {S}_u=\{(u_{1},u_{2},\ldots,u_{L})\in\mathbb{R}^r:\\
\label{Eq3_6} &&~~~~~~~~~~u_i~ \mbox{is a $r_i$-dimensional binary row vector},i=1,\ldots,L, r=\sum_{i=1}^Lr_i\},\\[3mm]
\label{Eq3_06}
P_{u_{1},u_{2},\ldots,u_{L}}&=&{\int_{\Xi_{(u_{1},u_{2},\ldots,u_{L})}}p(y_{1},y_{2},\ldots,y_{L}|\theta)}dy_{1}dy_{2}\ldots
dy_{L},\\[3mm]
\label{Eq3_006}
{\Xi_{(u_{1},u_{2},\ldots,u_{L})}}&=&\{(y_1,y_2,\ldots, y_L):
I_1(y_{1})=u_{1},I_2(y_{2})=u_{2},\ldots, I_L(y_{L})=u_{L}\}.
\end{eqnarray}
Note that $f_U(u_{1},u_{2},\ldots,u_{L}|\theta)$ is determined by
$p(y_{1},y_{2},\ldots,y_{L}|\theta)$ and sensor quantizers
$I_1(y_1)$, $\ldots$, $I_L(y_L)$.

Thus, the quantized observation population $U$ has a pmf parameter
family
$\{f_U(u_{1},u_{2},\ldots,u_{L}|\theta)\}_{\theta\in\Theta\subseteq\mathbb{R}^k}$
which yields the following log likelihood function of quantized
samples $\vec{U}$ by  (\ref{Eq3_5})-(\ref{Eq3_006}):
\begin{eqnarray}%
\label{Eq3_7}l(\theta|\vec{U})
&\triangleq&\log{\prod_{n=1}^Nf_U(U_{1n},U_{2n},\ldots,U_{Ln}|\theta)}\\[3mm]
\nonumber
&=&\sum_{n=1}^N\log{\int_{\{I_1(y_1)=U_{1n},\ldots,I_L(y_L)=U_{Ln}\}}}p(y_1,\ldots,y_L|\theta)dy_1\ldots
dy_L\\[3mm]
\label{Eq3_9} &=&\sum_{m=1}^{2^r}K_m\log{f_U(\vec{u}_m|\theta)}
\end{eqnarray}
where $K_m=\#\{(U_{1n},U_{2n},\ldots,U_{Ln})=\vec{u}_m\in S_u,
n=1,\ldots, N\}$, $\sum_{m=1}^{2^r}K_m=N$; $S_u$ is defined by
(\ref{Eq3_6}); $\#\{\cdot\}$ is the cardinality of the set. The
parameter vector $\theta$ is estimated by maximizing the log
likelihood function (\ref{Eq3_9}) or equivalently solving the
equation:
\begin{eqnarray}
\label{Eq3_10} &&\frac{\partial}{\partial\theta}l(\theta|\vec{U})=0,
\end{eqnarray}
whose solution is denoted by $\hat{\theta}$. In
\cite{Shen-Varshney-Zhu12}, we have considered the estimation
problem in detail and have presented the regularity conditions for
$p(y_{1},y_{2},\ldots,y_{L}|\theta)$  and quantizers $I(y|r)$ that
guarantee that $\hat{\theta}$ is asymptotically efficient.

\subsection{ Maximum likelihood estimation with feedback}
As indicated earlier, to improve the detection performance in the
distributed detection system, the quantizers  are updated and fed
back to the sensors for use at the following time. The quantizers
$I(y|r)$ defined in (\ref{Eq2_5}) used at a given time $j$,
$j=1,\ldots,J$, are known as one group of quantizers.
To distinguish different groups, we use superscript $^{(j)}$ and
change notations $n, N$ to $n_j, N_j$ respectively, $j=1,\ldots,J$.
Assume
that, for the $j$-th group of quantizers $I^{(j)}(y|r)$, $N_j$ joint
samples $\{(Y_{1n_j},\ldots,Y_{Ln_j})\}_{n_j=1}^{N_j}$ are observed
and the quantized observations are denoted by $\vec{U}^{(j)}$. The
corresponding observation population denoted by $U^{(j)}$ whose pmf
can be similarly defined by (\ref{Eq3_5}) and be denoted by
\begin{eqnarray}
\label{Eq3_37} f_U^{(j)}(u_{1},u_{2},\ldots,u_{L}|\theta),
j=1,\ldots,J.
\end{eqnarray}

Since the samples are temporally independent, we can estimate
$\theta$
 by maximizing the log likelihood function:
\begin{eqnarray}%
\label{Eq3_38} l(\theta|\vec{U}^{(1)},\ldots,\vec{U}^{(J)})
&=&\log{\prod_{j=1}^J\prod_{n_j=1}^{N_j}f_U^{(j)}(U_{1n_j}^{(j)},U_{2n_j}^{(j)},\ldots,U_{Ln_j}^{(j)}|\theta)}\\[3mm]
\nonumber&=&\sum_{j=1}^J\sum_{n_j=1}^{N_j}\log{f_U^{(j)}(U_{1n_j}^{(j)},U_{2n_j}^{(j)},\ldots,U_{Ln_j}^{(j)}|\theta)}\\[3mm]
\nonumber&=&\sum_{j=1}^J\sum_{n_j=1}^{N_j}\log{\int_{\{I_1^{(j)}(y_1)=U_{1n_j}^{(j)},\ldots,I_L^{(j)}(y_L)=U_{Ln_j}^{(j)}\}}}
p(y_1,\ldots,y_L|\theta)dy_1\ldots
dy_L\\
\label{Eq3_09}
&=&\sum_{j=1}^J\sum_{m=1}^{2^r}K_m^{(j)}\log{f_U^{(j)}(\vec{u}_m|\theta)}
\end{eqnarray}
where
$K_m^{(j)}=\#\{(U_{1n_j}^{(j)},U_{2n_j}^{(j)},\ldots,U_{Ln_j}^{(j)})=\vec{u}_m\in
S_u, n_j=1,\ldots, N_j\}$, $\sum_{m=1}^{2^r}K_m^{(j)}=N_j$; $S_u$ is
defined by (\ref{Eq3_6}); $\#\{\cdot\}$ is the cardinality of the
set. Equivalently, we solve the equation:
\begin{eqnarray}
\label{Eq3_41}&&\frac{\partial}{\partial\theta}l(\theta|\vec{U}^{(1)},\ldots,\vec{U}^{(J)})=0,
\end{eqnarray}
whose solution is denoted by $\hat{\theta}_{R}$. In
\cite{Shen-Varshney-Zhu12}, we have proved that $\hat{\theta}_{R}$
is an asymptotically efficient estimator with an asymptotic variance
equal to the inverse of a convex linear combination of Fisher
information matrices based on $J$ groups of different quantizers.
These results are summarized in the following Lemma.

\begin{lemma}\label{thm_2}
 There are $J$ groups of
different sensor quantizers $I^{(j)}(y|r)$, $j=1,\ldots,J$. Assume
that $p(y_{1},y_{2},\ldots,y_{L}|\theta)$ and quantizers
$I^{(j)}(y|r)$ generate the quantized observations and the quantized
pmf $f_U^{(j)}(u_{1},u_{2},\ldots,u_{L}|\theta)$ defined by
(\ref{Eq3_37}) satisfies the regularity conditions (C1)--(C7) in
\cite{Shen-Varshney-Zhu12}. The true parameter vector is denoted by
$\theta^*$. Then,
\begin{eqnarray}
\label{Eq3_42} \sqrt{N}(\hat{\theta}_{R}-\theta^*)\longrightarrow
N(0,\mathcal {I}^{-1}(\theta^*;I^{(1)}(\cdot),\ldots,I^{(J)}(\cdot))
\end{eqnarray}
where $N=\sum_{j=1}^JN_j, N_j\rightarrow\infty, j=1,\ldots,J,$
\begin{eqnarray}
\label{Eq3_43} \mathcal
{I}^{-1}(\theta^*;I^{(1)}(\cdot),\ldots,I^{(J)}(\cdot))&\triangleq&\left(\sum_{j=1}^J\frac{N_j}{N}\mathcal
{I}(\theta^*;I^{(j)}(\cdot))\right)^{-1}\\
\label{Eq3_44} &=&N\left(\sum_{j=1}^JN_j\mathcal
{I}(\theta^*;I^{(j)}(\cdot))\right)^{-1}.
\end{eqnarray}
$\left(\sum_{j=1}^JN_j\mathcal
{I}(\theta^*;I^{(j)}(\cdot))\right)^{-1}$ is the Cram\'{e}r-Rao
lower bound for $N$ quantized observations, where $\mathcal
{I}(\theta^*;I^{(j)}(\cdot))$ is the Fisher information matrix for
one quantized sample of $U^{(j)}$. That is, $\hat{\theta}_{R}$ is a
consistent and asymptotically efficient estimator of $\theta^*$.

\end{lemma}

\section{Distributed Detection System Design Using MLE with Feedback}\label{sec_4}
When the prior probabilities and conditional pdfs are unknown, the
basic idea of the distributed detection system design is as follows.
We begin with an initial set of quantizers at the sensors and send
quantized observations to the fusion center. The fusion center
starts with an initial fusion rule. Based on the received quantized
observations, the fusion center computes the MLE for the unknown
parameters, and obtains updated quantizers and the fusion rule.
Updated quantizers are fed back to the sensors and are used to
quantize the next set of observations. This iterative process is
continued several times to continually improve the parameter
estimates and thereby improving detection performance.
In summary, based on the MLE  with feedback and Algorithm 1 in
\cite{Shen-Zhu-He-You11} which is a near-optimal iterative algorithm
and can simultaneously design the quantizers and the fusion rule
when the prior probabilities and the conditionally dependent pdfs
are known, we have the following algorithm.

\begin{algorithm}[Distributed detection system design based on MLE with feedback]\label{alg_1}
~
\begin{description}
\item[Step 1:]~Initialize $L$ quantizers and the
fusion rule at first stage respectively, ~for $i=1,\ldots, L$,
\begin{eqnarray}
\label{Eq4_4} &&I_i^{(1)}(y_{im_i})=\textbf{0/1}, ~\mbox{for}~~m_i=1,\ldots, M_i, \\[3mm]
\label{Eq4_5} &&F^{(1)}(\vec{u}_j)=0/1, \mbox{for}~~\vec{u}_j\in S_u, j=1,\ldots, 2^L,
\end{eqnarray}
where $\textbf{0/1}$ is $r_i$ dimensional 0/1 vector; the
measurement space of the $i$-th sensor is discretized to $M_i$
regions; $S_u$ are defined by (\ref{Eq3_6}). $N_1$ samples are
sequentially observed and quantized to binary samples which are sent
to the fusion center. Let $t=1$ and $J=1$, go to next step.

\item[Step 2:]~Estimate parameter $\theta$ at the $t$-th stage:
the MLE with feedback $\hat{\theta}_R$ is computed based on all
previous quantized observations by maximizing (\ref{Eq3_09}).
Thus, we have $p(y_1,y_2,\ldots,y_L|\hat{\theta}_R)$, go to next
step.

\item[Step 3:]~Design sensor
quantizers and the fusion rule at the $t$-th stage: Based on
$p(y_1,y_2,\ldots,y_L|\hat{\theta}_R)$ and Algorithm 1 in
\cite{Shen-Zhu-He-You11}, sensor quantizers and the fusion rule are
iteratively searched for better detection performance until the
termination criterion of Algorithm 1 in \cite{Shen-Zhu-He-You11} is
satisfied. Thus, we have the  quantizers $I_i^{(t)}(\cdot),
i=1,\ldots, L$ and the fusion rule $F^{(t)}(\cdot)$, go to next
step.

\item[Step 4:]~Feedback:  If $t\leq T$ (T is the
upper bound on the number of feedbacks), let $t=t+1$ and set $J=t$.
The fusion center sends
the current quantizers $I_i^{(t-1)}(\cdot), i=1,\ldots, L$ to the
sensors, and then the sensors sequentially observe $N_t$ quantized
samples based on the current quantizers and send them to the fusion
center. Go to step 2. If $t>T$,  stop and the last quantizers are
transmitted to the sensors.
\end{description}
\end{algorithm}

In feedback step 4, communication can be reduced by only
transmitting the changes of the quantizers between two iteration
steps to sensors.

\section{Numerical Examples}\label{sec_5}


Let us consider a binary hypothesis testing problem with two sensors
\begin{eqnarray}
\nonumber&&H_0: Y_1\sim Gamma(3,4) ~~~~~ Y_2\sim Gamma(5,4),\\[3mm]
&&~~~~~~
c(v_1,v_2|H_0)\equiv1 ~~(\mbox{independent copula density}),\\[4mm]
\nonumber&&H_1: Y_1\sim Gamma(5,4) ~~~~~ Y_2\sim Gamma(7,4),\\[3mm]
&&~~~~~~
c(v_1,v_2|H_1,\theta_1)=(1+\theta_1)v_1^{-1-\theta_1}v_2^{-1-\theta_1}
(-1+v_1^{-\theta_1}+v_2^{-\theta_1})^{-2-1/\theta_1},\theta_1\in[-1,\infty)\backslash\{0\},
\end{eqnarray}
where the prior probability  $P_1$  and dependence parameter
$\theta_1$ of hypothesis $H_1$ are unknown and are required to be
estimated. We denote by $\theta\triangleq[P_1, \theta_1]$. Here, we
assume that the joint pdf under $H_0$ is independent and marginal
pdfs are known. $c(v_1,v_2|H_1,\theta_1)$ is the Clayton copula
density (see e.g. \cite{Nelsen99}) which is a frequently used copula
to model dependence. The actual value of $\theta$ (ground truth) is
that $P_1=0.2$ and $\theta_1=[ 0.5109, 1.0759, 2.1316]$ which
corresponds to Spearman's dependence measure  $\rho=[0.3,0.5,0.7]$
\footnote{Spearman's $|\rho|\leq1$ is a commonly used dependence
measure (see \cite{Nelsen99}).}. There is a one to one relationship
between $\theta_1$ and $\rho$ (see \cite{Nelsen99}).

The initial values of the quantizers chosen are
$I_1(y_1)=I[3y_1-60]$, $I_2(y_2)=I[-3y_2+60]$, and the initial
fusion rule used is the OR fusion rule.  For numerical computation,
we take a discretization step-size $\Delta=0.5$, $y_i\in[0,60]$. We
denote the probability of a false alarm and the probability of
detection by $P_f$ and $P_d$ respectively.

\begin{figure}[h]
\vbox to 9cm{\vfill \hbox to \hsize{\hfill
\scalebox{0.5}[0.5]{\includegraphics{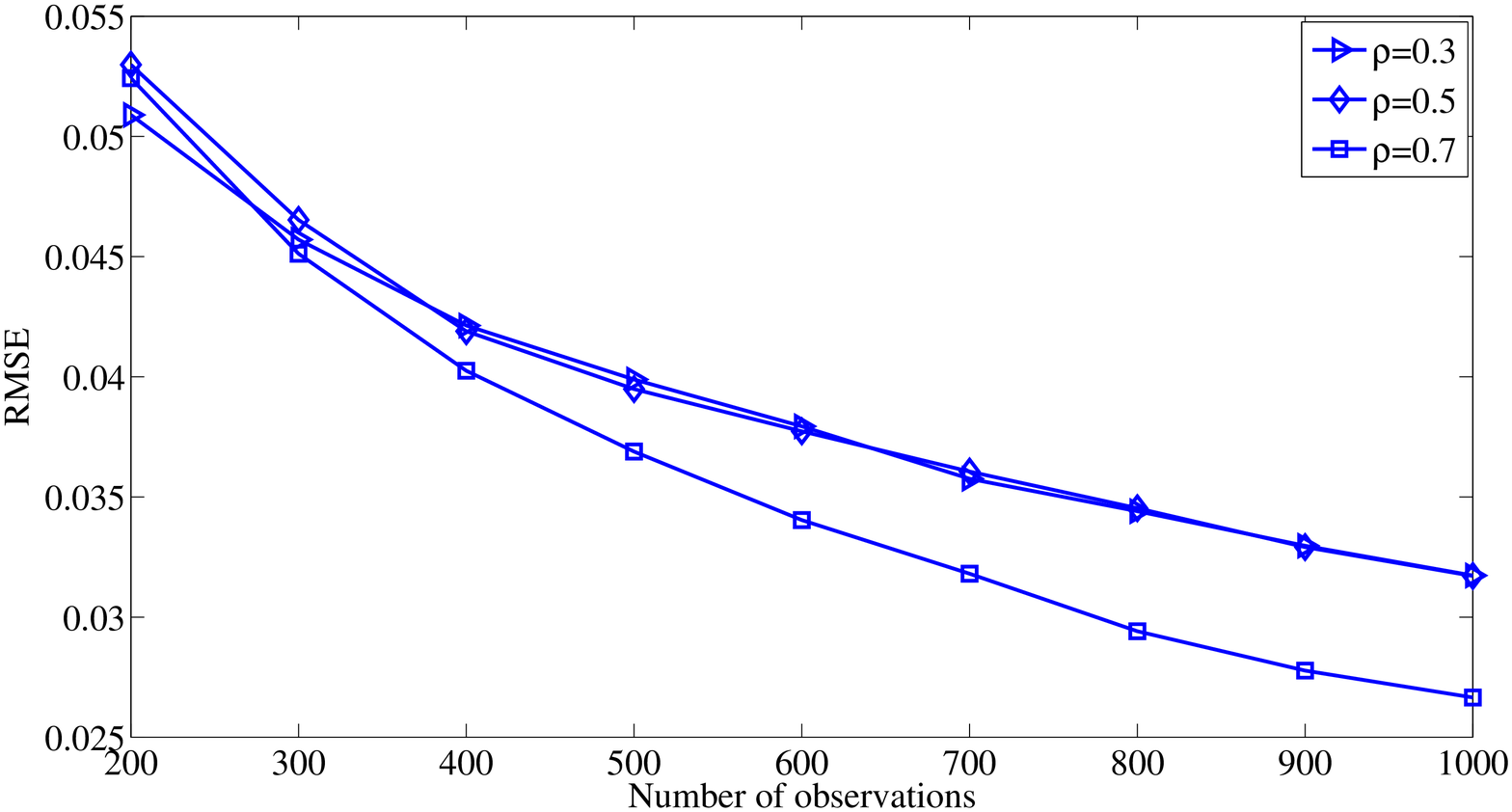}}
\hfill}\vfill}
\caption{RMSEs of the prior probability $P_1$ based on 5000 Monte
Carlo runs  for the cases of the dependence measure Spearman's
$\rho=[0.3,0.5,0.7]$.}\label{fig_1}
\end{figure}

\begin{figure}[h]
\vbox to 9cm{\vfill \hbox to \hsize{\hfill
\scalebox{0.5}[0.5]{\includegraphics{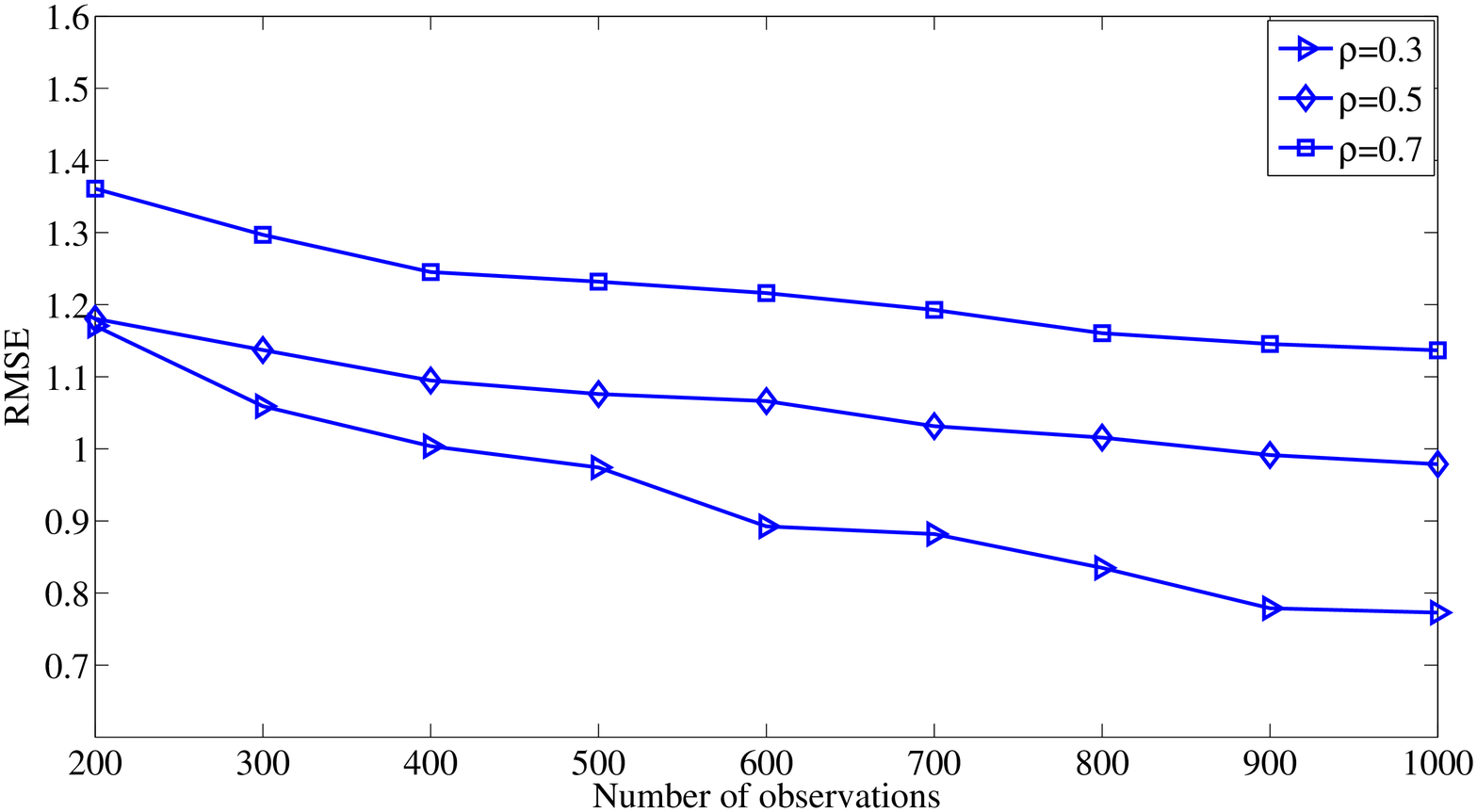}}
\hfill}\vfill}
\caption{RMSEs of the dependence parameter $\theta_1$ based on 5000
Monte Carlo runs for the cases of the dependence measure Spearman's
$\rho=[0.3,0.5,0.7]$}\label{fig_2}
\end{figure}

In Figure \ref{fig_1}, RMSEs of the MLE with feedback for the prior
probability $P_1$ are plotted based on 5000 Monte Carlo runs. Three
dependence cases with Spearman's $\rho=[0.3,0.5,0.7]$ are
considered. The feedback times increase from $J=2$ to $J=10$ and
each estimation step is done after receiving $100$ quantized
observations so that the number of observations becomes from
$J\times 100=200$ to $1000$. The cost parameters $C_{00}=C_{11}=0$
and $C_{10}=2$, $C_{01}=1$ are used in cost function Eq.
(\ref{Eq4_3}). Similarly, RMSEs of the MLE with feedback for the
dependence parameter $\theta_1$ are plotted in Figure \ref{fig_2}.

\begin{figure}[h]
\vbox to 9cm{\vfill \hbox to \hsize{\hfill
\scalebox{0.5}[0.5]{\includegraphics{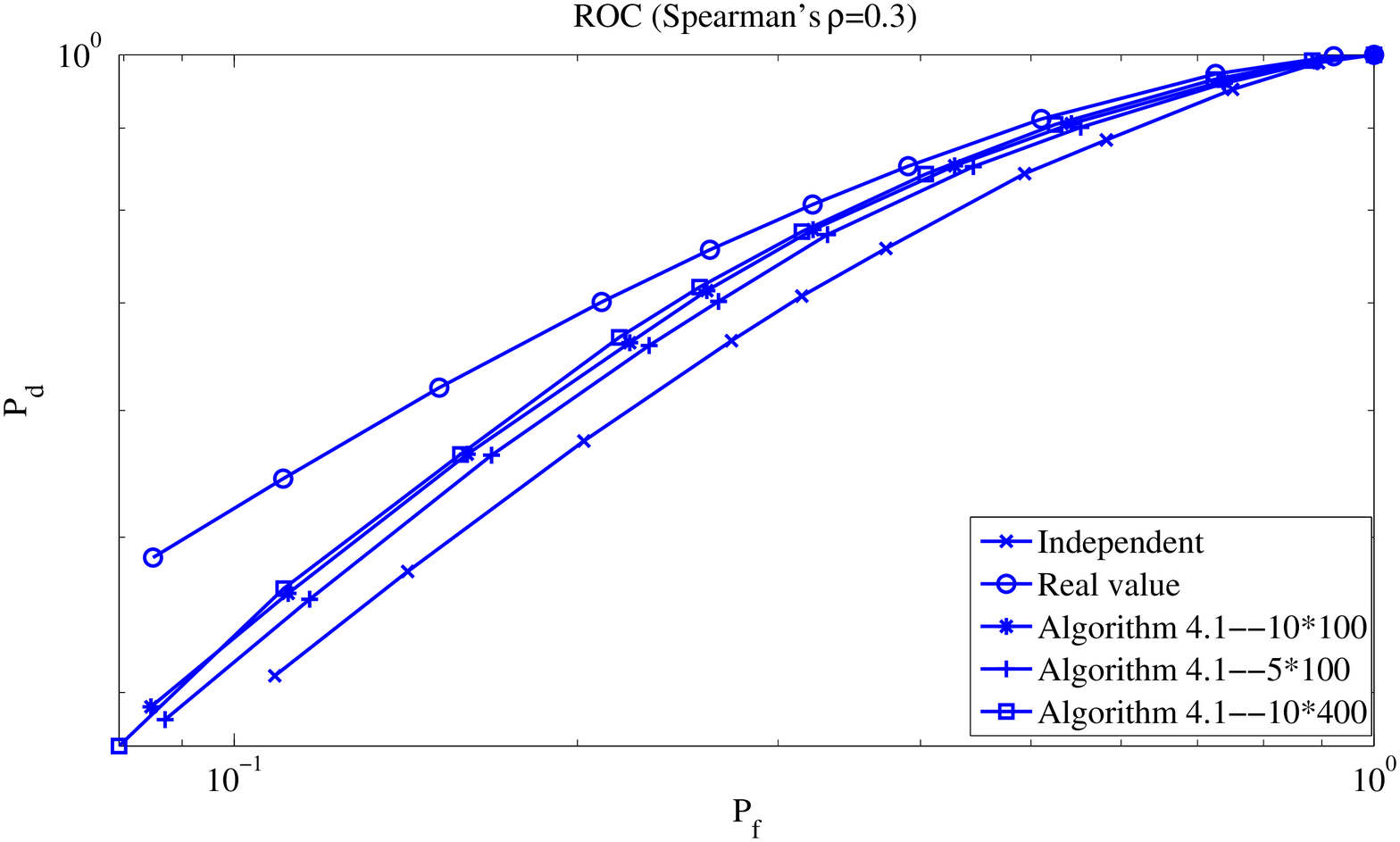}}
\hfill}\vfill}
\caption{Comparison of ROCs for three cases with Spearman's
$\rho=0.3$.}\label{fig_3}
\end{figure}

\begin{figure}[h]
\vbox to 9cm{\vfill \hbox to \hsize{\hfill
\scalebox{0.5}[0.5]{\includegraphics{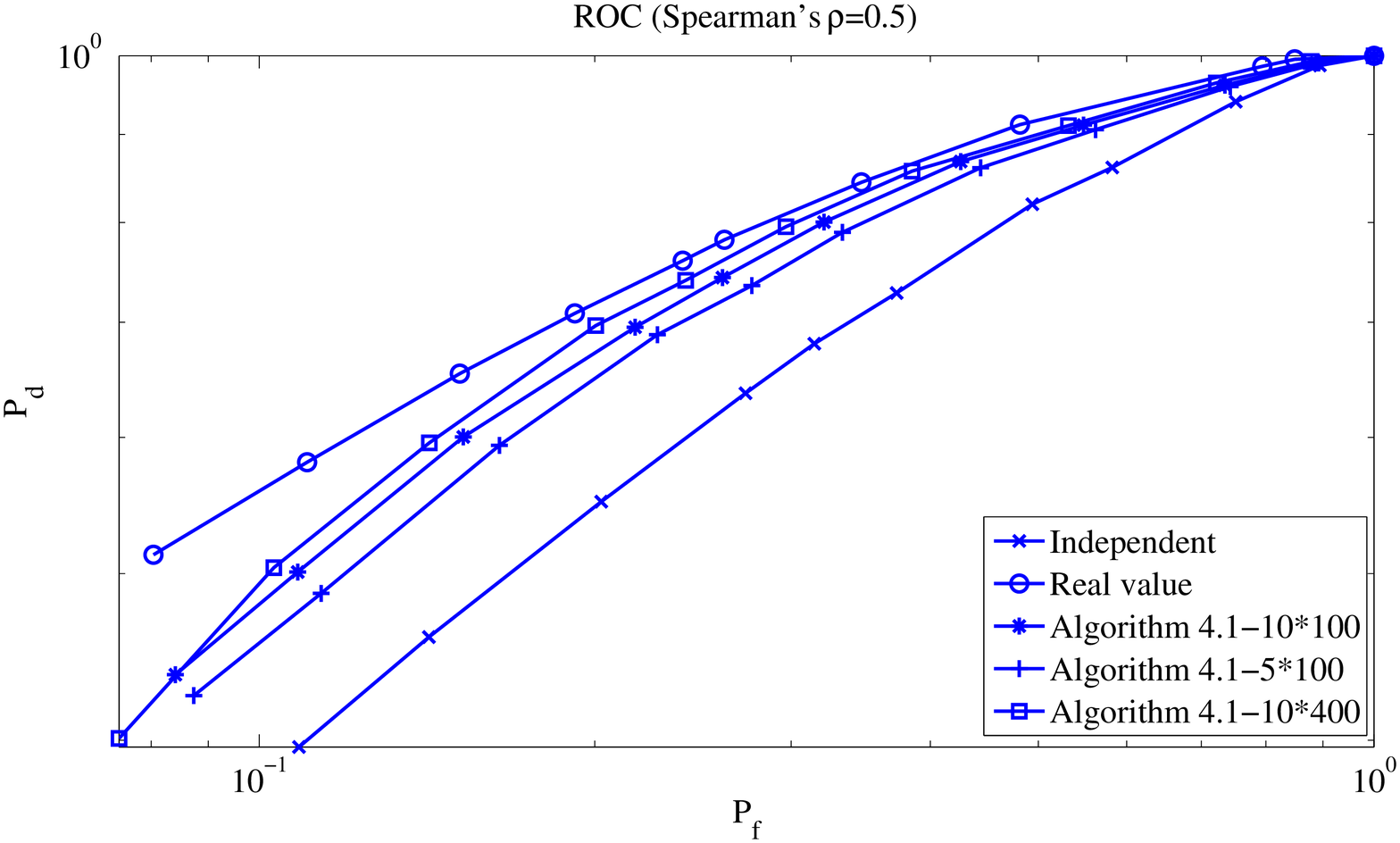}}
\hfill}\vfill}
\caption{Comparison of ROCs for three cases with Spearman's
$\rho=0.5$.}\label{fig_4}
\end{figure}

\begin{figure}[h]
\vbox to 9cm{\vfill \hbox to \hsize{\hfill
\scalebox{0.5}[0.5]{\includegraphics{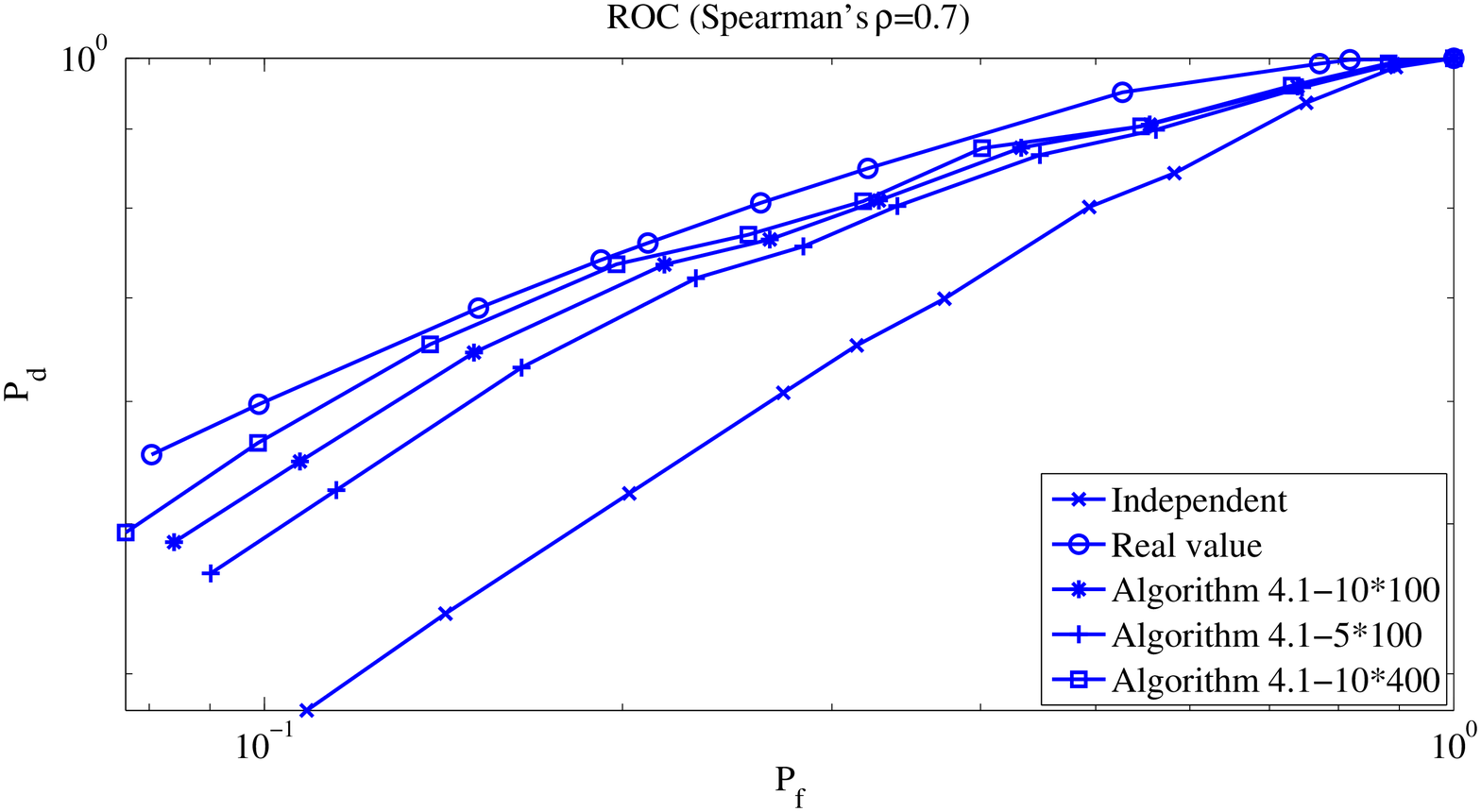}}
\hfill}\vfill}
\caption{Comparison of ROCs for three cases with Spearman's
$\rho=0.7$.}\label{fig_5}
\end{figure}

In Figures \ref{fig_3}--\ref{fig_5}, to evaluate the detection
performance of Algorithm \ref{alg_1},  the average ROC curves based
on 500 Monte Carlo runs are compared  for the following three cases:
1) Algorithm \ref{alg_1} based on MLE with feedback, where
``Algorithm \ref{alg_1}--10*100" means the number of feedbacks
$J=10$ and the number of observations in each estimation step
$N_j=100$, $j=1,\ldots,J$.
2) Assume independence under $H_1$ so that the joint density is the
product of marginals and with known prior probabilities.
3) The case with known  parameter values (the clairvoyant case). In
each Monte Carlo run, 400 test observations from $H_1$ and 1600 test
observations from $H_0$ are generated so that $P_f$ and $P_d$ can be
computed respectively. To plot several points on the ROCs, the cost
parameters $C_{00}=C_{11}=0$ and $C_{10}=2$, $C_{01}=2\times[ 0.6,~
0.5, ~0.4, ~0.3,~ 0.25,~ 0.2,~ 0.15, ~0.1, ~0.05,~ 0.02]$ are used
in cost function Eq. (\ref{Eq4_3}).

From Figures \ref{fig_1}--\ref{fig_5}, we have the following
observations:
\begin{enumerate}
\item From Figures \ref{fig_1} and \ref{fig_2},  RMSEs  decrease as
the number of observations increases. However, the rate of decay
becomes slow, especially in Figure \ref{fig_2}. Thus, to improve
performance at a later stage, more feedback times and samples are
required.


\item RMSE of the parameter $P_1$ in the case of $\rho=0.7$ is the smallest
among the three cases in Figure \ref{fig_1}. However,  RMSE of the
parameter $\theta_1$ in the case of $\rho=0.7$ is largest in Figure
\ref{fig_2}. The reason may be that the estimated pdf in this case
is closer to the actual pdf. In addition, from Figure \ref{fig_2},
RMSE of $\theta_1$ for $\rho=0.3$ is less than those for $\rho=0.5$
and $\rho=0.7$. The reason may be that the value of $\theta_1$
corresponding to $\rho=0.3$ is less than those corresponding to
$\rho=0.5$ and $\rho=0.7$.

\item From Figures \ref{fig_3}--\ref{fig_5},  the new  detection  Algorithm \ref{alg_1} based on MLE
with feedback is much better than the usual detection method based
on the assumption of independent observations. For fixed number of
observations in each estimation step, the better performance can be
obtained by increasing the  number of feedbacks from 5 to 10. For
fixed the number of feedbacks 10, better performance can be obtained
by increasing the number of observations in each estimation step,
especially for the case of the larger dependence parameter
$\rho=0.7$.


\end{enumerate}

\section{Conclusion}\label{sec_6}
In this paper, distributed Bayesian detection problems with unknown
prior probabilities of hypotheses and unknown conditional pdfs have
been considered. The conditional joint pdf was represented in a
parametric form by using the copula framework. The unknown
parameters included dependence parameters and marginal parameters.
MLE with feedback based on quantized data has been proposed to
estimate the unknown parameters. Its asymptotic efficiency can be
guaranteed by employing the result that we have developed in
\cite{Shen-Varshney-Zhu12} with an asymptotic variance which is
equal to the inverse of a convex linear combination of Fisher
information matrices based on $J$ groups of different feedback
quantizers.
These estimates were iteratively used to refine the quantizers and
the fusion rule to improve distributed detection performance by
using feedback. Numerical examples show that the new detection
method based on MLE with feedback is much better than the usual
detection method based on the assumption of conditionally
independent observations. Better detection performance can be
obtained by increasing the number of feedbacks and the number of
observations in each estimation step.
%
%

Future work will involve distributed detection and distributed
location estimation of non-ideal  systems where sensor outputs are
to be communicated through noisy, possibly coupled channels to the
fusion center.

\section*{Acknowledgment}
We would like to thank Hao He and Arun Subramanian for their
suggestions on simulations of this paper.



%

\end{document}